\documentstyle[12pt]{article}
\def\jump{\vskip0.5truecm}

\def\be{\begin{equation}}
\def\ee{\end{equation}}
\def\l{\label}
\def\ba{\begin{array}}
\def\ea{\end{array}}

\def\V{{V}}

\def\refe#1{(\ref{#1})}

\def\ltap{\ \raisebox{-.4ex}{\rlap{$\sim$}} \raisebox{.4ex}{$<$}\ }
\def\gtap{\ \raisebox{-.4ex}{\rlap{$\sim$}} \raisebox{.4ex}{$>$}\ }

\def\npb#1#2#3{    {\it Nucl. Phys. }{\bf B #1} (19#2) #3}

\def\plb#1#2#3{    {\it Phys. Lett. }{\bf B #1} (19#2) #3}
\def\prd#1#2#3{    {\it Phys. Rev. }{\bf D #1} (19#2) #3}

\def\prl#1#2#3{    {\it Phys. Rev. Lett. }{\bf #1} (19#2) #3}

\def\mpla#1#2#3{   {\it Mod. Phys. Lett. }{\bf A #1} (19#2) #3}

\begin{document}
\begin{titlepage}
\vspace*{-1.5cm}
\begin{center}

\hfill IC/96/16
\\[1ex]  \hfill February, 1996

\vspace{5ex}
{\Large \bf Upper bound on all products
\\[-0.5ex] of R-parity violating couplings $\lambda'$ and $\lambda''$ 
\\[1ex] from proton decay}

\vspace{3ex}
{\bf
Alexei Yu. Smirnov$^{a,b}$ and Francesco Vissani$^{a,c}$
}

{\it \vspace{1ex} ${}^a$ International Centre for Theoretical Physics,
ICTP \\[-1ex] Via Costiera 11, I-34013 Trieste, Italy
}

{\it \vspace{1ex} ${}^b$ Institute for Nuclear Research, \\[-1ex] Russian
Academy of Sciences, \\[-1ex] 117312 Moscow, Russia
}

{\it
\vspace{1ex}  ${}^c$ Istituto Nazionale di Fisica Nucleare, INFN
\\[-1ex] Sezione di Trieste
}

\vspace{6ex}
{ABSTRACT}
\end{center}

\begin{quotation}

We prove that {\em any} product 
of R-parity violating couplings 
$\lambda'$ (L-violating) and $\lambda''$ (B-violating) 
can be strongly restricted by proton decay data. 
For any pair $\lambda'$ and $\lambda''$  the 
decay exists at least at one loop level. 
For squark masses below 1 TeV 
we find the conservative bounds 
$|\lambda' \cdot \lambda''| < 10^{-9}$
in absence of squark flavor mixing,
and $|\lambda' \cdot \lambda''| < 10^{-11}$
when this mixing is taken into account.
We study the dependence of the bounds
on the flavor basis in which R-parity breaking couplings 
are determined.

\end{quotation}
\end{titlepage}
\vfill\eject

1. The proton decay gives very strong bounds on 
the products of the R-parity 
violating couplings 
$\lambda'$ (L-violating) 
and $\lambda''$ (B-violating) 
involving {\em light} generations \cite{Weinberg}.
The decay takes place at tree level, and  
for squark masses about 1 TeV one gets \cite{Hinchliffe}: 
\be
|\lambda'\cdot \lambda''|\ltap 10^{-24} . 
\l{tree}
\ee
What are the bounds on the couplings 
involving heavy generations? 
Whether there are unrestricted couplings? These 
questions are  important not only for 
phenomenology \cite{p1,p2,p3,p4,p5,p6,p7} 
but also for understanding the
origin of the R-parity, as well as  for 
the unification of interactions. 
Notice, for instance, that in 
stringy unified models the R-parity violation could be the only 
mechanism of the proton decay. 

In the context of the Grand Unified Theories it was
shown  that the quark-lepton symmetry 
usually leads to separate strong bounds on 
$\lambda'$  and $\lambda''$ \cite{Smirnov}. 
The situation can be different in absence of 
Grand Unification. 
It was claimed that certain products of
$\lambda'$  and $\lambda''$ 
constants  may be large without inducing proton 
decay at observable level \cite{Carlson}. 

In this Letter we prove that 
operators relevant to the proton decay
are always present in the 
one loop effective lagrangian, 
and they imply strong bounds 
on any  product of the couplings 
$\lambda' \cdot \lambda''$.\jump

2.  Let us introduce the 
L-violating coupling  constant
$\lambda'_{ijk}$  and the B - violating coupling 
constant $\lambda''_{mnp}$,  
where $i,j,k,m,n,p=1,2,3$ are the generation indices, 
as the constants of the R-parity violating interactions:
\be
\begin{array}{rl}
 & \lambda_{ijk}'\ D_i^{c\alpha}\ 
       (\nu_j\ S^d_{kl}D_l^\alpha-E_j\ 
S^u_{kl}
\ U_l^\alpha)\\
  +&  \lambda_{mnp}''\ \epsilon_{\alpha\beta\gamma}\
                       D_m^{c\alpha}\  D_n^{c\beta}\ U_p^{c\gamma} 
\end{array}
\l{R-parity-violating}
\ee
which are consistent with the Standard Model symmetry. 
Here, $E_i,\nu_i,D^c_i,D_i,U^c_i,U_i$ are the superfields
with charged leptons, neutrinos, down- and up-type-quarks,
and $\alpha,\beta,\gamma$ are color indices.
Notice that in (\ref{R-parity-violating})  
the sum is over the flavor index $l$ only. 
The couplings  are written in terms of superfields 
whose fermionic components coincide with the mass
eigenstates (in all orders of  perturbation theory). 
The unitary matrices $S^d$ and 
$S^u$ connect the fermionic states of 
the original basis in which the couplings are defined
with the mass states. The product 
\be
{S^u}^\dagger  S^{d} = \V
\l{CKM}
\ee 
gives the Cabibbo-Kobayashi-Maskawa matrix  $\V $.

Let us stress that the R-parity violating couplings
depend on separate rotations of the upper, 
$S^u,$ and down, $S^d,$ components of the quark doublet.  
Correspondingly, the bounds on the couplings will depend on the 
basis in which they are defined. 
One can also introduce flavor rotations $S$ for the 
other quark and lepton superfields.

The scheme of the proof that proton decay is induced by 
any pair of the couplings $\lambda',  \lambda''$ is the 
following. 
Using an arbitrary pair $\lambda'$ and  $\lambda'',$ 
one can construct 4-field effective operators 
with (B - L) or (B + L) 
violation. If only light quarks and leptons are involved, 
these operators leads to the 
proton decay already at tree level. If the operator contains 
heavy quarks the proton decay will be forbidden kinematically. 
However, additional interactions with charged Higgs bosons 
(or Higgsinos, $W$-bosons or Wino), which violate flavor 
and thus can transform heavy quarks into  
light ones, lead to operators inducing proton decay. 
Furthermore we will show that for all pairs of couplings 
such an operator appears at the one loop level.

The coupling of the physical charged Higgs boson $h^+$ 
with the quarks and the squarks can be written as: 
\be
\ba{rl}
V_{kl}\ h^+
\left\{\displaystyle\ 
\frac{m_{d_l}}{v} \right. &
\left[ 
\tan\beta\cdot {u_k} {d^c_l} 
- (\mu+A_{d_l}\tan\beta)\cdot \tilde u_k {{\tilde{d}}^c_l}
\right]^*  \\[2ex]
 \displaystyle
+\ 
\frac{m_{u_k}}{v} & \left.
\left[ 
\cot\beta\cdot {u^c_k} {d_l} 
- (\mu+A_{u_k}\cot\beta)\cdot {{\tilde{u}}^c_k} \tilde d_l
\right]\ 
\right\} + {\rm h.c.}.
\ea
\l{higgs}
\ee
The tilded fields correspond to the scalar quarks,
$A_{u,d}$ are the soft breaking parameters, $\mu$ 
is the  supersymmetric mass, $\tan\beta$
is the ratio of vacuum expectation values of the two 
Higgs fields and $v=$ 174 GeV.
Notice that 
the up-quark masses,  
$m_{u_k}$,  
appear in the interactions of 
$u^c_k$ and $\tilde{u}^c_k$ fields, 
whereas the interactions of the $d^c_l$
and $\tilde{d}^c_l$ are proportional to the down-quark masses
$m_{d_l}$.

The squarks $\tilde d_n$ and $\tilde d^c_n$  mix by the mass:
\be
m^2_{\tilde d_n }({\rm \scriptstyle LR}) = 
m_{d_n} (A_{d_n} + \mu \tan \beta),
\l{LR-mixing}
\ee
and similarly will do the $u$-type-squarks.\jump

3. In general interfamily connections enhance the 
proton decay. Therefore to get the most conservative 
bound one should use the basis with   
maximally suppressed  connection.  
For this purpose 
we choose  the basis in which 
\be 
S^d = I, \ \ \  S^u = {\V}^\dagger \ .
\ee
In the proof we will use the L-violating coupling
with neutrino only (the first term in 
\refe{R-parity-violating}) which does not contain 
any mixing at all. 
We will discuss the dependence of the result on the basis 
in Sect.\ 6.

The generation structure of $\lambda'$
and $\lambda''$ plays a crucial role.
Let us recall in this connection that, according to
Eq.\ \refe{R-parity-violating},
in the $\lambda'$-coupling we use 
the index $i$ to denote the generation of $D^c,$
the index $j$ for the lepton doublet
and $k$ for the quark doublet.
In the $\lambda''$-coupling the indices $m$ and $n$ are prescribed to
the $D^c$ superfields, whereas the index $p$ is for the 
$U^c$ superfield. 

All possible pairs of couplings 
$\lambda'_{ijk}$, $\lambda''_{mnp}$ can be divided into two 
classes: \vskip-\parskip
\noindent $(i)$ Pairs with ``matching'' of the $D$-fields, when   the
generation index of the  $D$ or $D^c$ 
field from the lepton-violating vertex  coincides with the 
generation index 
of the $D^c$ field from the baryon-violating vertex. 
That is $i = m$ or $i = n$ (``$D^c D^c$-matching"),  
or      $k = m$ or $k = n$  (``$D^c D$-matching").\vskip-\parskip
\noindent $(ii)$ Pairs without ``matching'' of the $D$-fields. 
Taking into account the antisymmetry 
of $\lambda_{m n p}''$ in $m$ and $n$ one finds that 
``no-matching'' case is realized only if $i = k$ and 
$i, m, n$ are all different, {\em i.e.}\ $D^c$-fields of 
all three generations should be present. 
Let us construct and estimate the diagrams for these  
two cases.

$(i)$ Pairs with ``matching''.

Suppose first that there is the       
``$D^c D^c$-matching" and take $i = m$  
for definiteness. 
We can connect the B- and the L-violating 
vertices by one $\tilde d^c$-propagator.    
In the obtained  tree level diagram (Fig.\  1a)   
at most one external line  is $b$ or $b^c$. 
(If both of them are $b,$ then one has situation  with 
``$D^c D$-matching" which will be considered later).    
The case $d^c_n \equiv b^c$  
corresponds to $d_k = d$ or $s;$
the former quark should be connected with $u^c_p$ by charged Higgs,
whereas the latter can be an external line of an operator
which gives proton decay.
In such a way one gets the vertex type diagram  shown Fig.\  1c. 
Emitting $h^+,$ $b^c$ transforms into an $u$-quark, 
and, absorbing $h^+,$ $u^c_p$ transforms into a $d$ 
(or a $s$) quark. The resulting effective operators,  
$u d d \nu$ or  $u d s \nu$,  contain all  light fields. 
Using the coupling constants of the Higgs 
(\ref{higgs}) we find 
the suppression factor,  $\xi$, of the one loop 
diagram with respect to 
the tree level diagram $\xi \equiv (loop)/(tree)$:  
\be
\xi  \approx
\frac{1}{(4 \pi)^2} \times 
\frac{m_{u_p}}{v} \
\frac{m_{b}}{v} \times
V_{13}\
V_{p1}\ . 
\label{suppvert}
\ee

If $d_k \equiv b$, then one should connect by Higgs exchange  
$u^c_p$ 
with $d_k$, so that $d_k \rightarrow u$ and 
$u^c_p \rightarrow d$ or $s$. 
The resulting box diagram, of the type shown in Fig.\ 1e, 
has a suppression factor  similar 
to that in (\ref{suppvert})  with substitution 
$m_{b} V_{13} \rightarrow m_{u} V_{1k}$. 

Suppose that in the  
tree-level diagram both $d$-lines are not $b;$
the decay is allowed unless $u^c_p$ are $c^c$- or $t^c$-quarks, 
or if both $d$-lines are $s$-quarks.
In both cases, the Higgs exchange between
a $d^c$ line and $u^c_p$ leads to the proton decay 
at one loop level, and the suppression 
factors are similar to that in Eq.\ (\ref{suppvert}). 
 
In the case of ``$D^c D$-matching" among the 
external lines in the diagram (Fig.\ 1b) 
at most one is $b$ and at least 
one coincides with $d$ or $s$. Similarly to the previous case 
one should connect this heaviest $d^c$-quark line with 
$u_p^c,$ thus arriving at vertex (Fig.\  1d) or box (Fig.\  1f) 
diagrams. The suppression factor equals
\be
\xi \approx 
\frac{1}{(4 \pi)^2} \times 
\frac{m_{d_n}}{v} \
\frac{m_{u_p}}{v} 
\frac{m_{b}}{\tilde{m}} \times
V_{1n}\ 
V_{1p}\ ,
\label{supp+}
\ee
where $\tilde{m}$ is the typical 
squark mass and the factor $m_{b}/{\tilde{m}}$ 
follows from the mixing of the left and right squarks 
(\ref{LR-mixing}). 
Notice that the diagrams shown 
lead to (B+L) conserving operators
$u d d \bar \nu$,   $u d s \bar \nu$.

$(ii)$ ``No-matching" case. 

Recall that in ``no-matching'' case the
three $D^c$ should all be of different 
flavors and therefore one 
of them is $B^c$ and two others are $D^c$ and $S^c$. 
In this case one can always construct 
the  diagram of the type shown in Fig.\ 2. 
The $\tilde u^c_p$-squark from 
the baryon-violating vertex can emit 
an Higgs boson and a $\tilde d_k$-squark,  
and the latter can be adsorbed by the
lepton-violating vertex. 
The higgs field in turn is absorbed by   
the $d^c_i \equiv b^c$ line, so that $b^c \rightarrow u$  
(Fig.\  2a, 2b). 
As a result one gets the operator 
$u d s  \nu$. Integration over the 
loop, and the Lorentz structure of the vertices,  pick 
up the momentum of an external quark $p_q \sim m_N/3$,  
where $m_N$  is the nucleon mass. The suppression factor 
can be estimated as 
\be
\xi \approx 
\frac{1}{(4 \pi)^2} \times 
\frac{m_b}{v}\ 
\frac{m_{u_p}}{v}\ 
\frac{p_q}{\tilde{m}} \times
V_{31}\ 
V_{pk} \ 
\tan\beta
\label{suppnon-m}. 
\ee
For a given couplings one can construct 
another version of the diagram:  
the field $b^c$ changes 
chirality, $b^c \rightarrow b,$ and $b$  transforms into $u^c$
by absorbing an Higgs field (Fig.\  2c, 2d). 
In this case 
\be
\xi \approx 
\frac{1}{(4 \pi)^2} \times 
\frac{m_u}{v} \
\frac{m_{u_p}}{v} \
\frac{m_b}{\tilde{m}} \times 
V_{31}\
V_{pk} \ 
\tan\beta
\label{suppnon-m1}. 
\ee
(Notice that now $p_q/\tilde{m}$ is substituted by 
$m_u/v$). The flip of chirality may take place in 
$\tilde d-$ (Fig.\  2e) or $\tilde u-$ (Fig.\  2f) 
propagators connecting 
the B- and the L-vertices. The suppression factors are 
of the same type as in (\ref{suppnon-m1}). 

Thus we have shown that for any pair 
$\lambda'$, $\lambda''$ there is  (at least one) one loop diagram 
which leads to the proton decay. 
Usually for each pair one can find several diagrams, 
and moreover in some cases, as we will see in Sect.\ 4,
new diagrams give even bigger contributions.

Diagrams similar to those in Figs.\ 1, 2 
arise due to the $W$-boson, 
the would-be Goldstone boson and the charginos 
(mixed states of charged Higgsinos and Wino) 
exchanges, instead of the Higgs exchange.  
Effective tree level or one loop operators leading to the emission 
of light charged leptons are possible for certain
pairs of couplings in \refe{R-parity-violating}.
Additional sources of flavor violation 
could be related to the interactions of gluinos and neutralinos. 

For each channel of the proton decay 
one can find different contributions. 
Since in general they have different Lorenz structure and
depend on different unrelated parameters (masses {\em etc.}),   
we suggest that there is no accidental strong cancellation 
of different contributions.\jump

4. Let us find now a conservative bound on the 
product of the R-parity violating couplings and 
identify the pairs of coupling which are less restricted.  
For this  we evaluate the suppression factors $\xi$ 
corresponding  to each constructed diagrams. 
We assume squark masses around 1 TeV,
and take quark running masses 
at the squark scale. 
Since for large values of $\tan\beta$ 
the suppression is weaker, we will use $\tan\beta=2,$
compatible with the top Yukawa coupling being
perturbative up to the Planck scale.

Considering the  
diagrams Fig.\  1,2 only,  we found that the 
smallest factor $\xi$ is for the pair of couplings 
\be 
\lambda'_{3j3}\ \lambda''_{121} 
\ee
(the L-violating and the B-violating vertices contains
$B^c \nu_j B$ and $D^c S^c U$ correspondingly). 
There is no ``$D$-matching''. Dominating contribution comes 
from ``no-matching" diagram of Fig.\  2a. The suppression factor 
is given in (\ref{suppnon-m}) with $V_{pk} = V_{13}$: 
$\xi \approx 10^{-17}$. The factor results in   rather weak  bound 
on the couplings  
$|\lambda'_{3j3} \cdot \lambda''_{121}| < 10^{-7}$. 
However for these couplings another type of diagrams exists, 
the one with neutrino-Zino mixing,  
which  leads to a much stronger bound.  

The neutrino and the Zino are mixed 
by one loop diagrams generated, {\em e.g.},\ 
by the R-parity violating coupling $\lambda'_{3j3}\ b^c \nu \tilde b $, 
the gauge interaction  $g\ \tilde{b}^* b \tilde Z$ and
by  the mass term $m_b\ b b^c$. The 
Zino couples with one of the squarks 
emitted from the B-violating vertex thus leading to the proton decay. 
Notice that for the existence of such a diagram it is 
important that all the quarks in the B-violating vertex are light. 
The suppression factor for the diagram with 
$\nu - \tilde Z$ mixing can be estimated as 
\be 
\xi \approx 
\frac{g^2}{(4 \pi)^2} \times 
\frac{m_b}{m_{\tilde Z}} 
\approx 7\cdot 10^{-6}. 
\label{nuzino} 
\ee
It gives the bound on the product of couplings 
of the order $10^{-18}$. 

Next weakest bound  
is for the product 
\be \lambda_{2j2}' \lambda_{131}''. 
\l{best}
\ee
For these couplings there is no proton decay
diagram with $\nu - \tilde Z$ mixing. 
The dominating contribution comes from 
the ``no-matching" diagram of Fig.\  2b, 
and the suppression factor is:
\be
\xi \approx 10^{-15} \ .
\label{less}
\ee
Other products, 
\be
\lambda'_{1j2}\ \lambda''_{231}, \ \ 
\lambda'_{3j2}\ \lambda''_{121}, \ \
\lambda'_{1j1}\ \lambda''_{231}, \ \
\lambda'_{2j1}\ \lambda''_{131}, \ \
\lambda'_{3j1}\ \lambda''_{121},
\label{next}
\ee
have a few times larger $\xi$ factors.
All but the third pair in \refe{next} correspond to the case 
of ``$D D^c$-matching.'' but since the ``matching'' is suppressed 
by the LR mixing the less suppressed diagrams
are again those of Fig.\ 2.

We conclude, using \refe{less}, that the conservative bound on {\em any}
product of $\lambda'$- and $\lambda''$-type coupling is:
\be
|\lambda'\cdot \lambda''|\ltap 10^{-9}\ .
\l{conservative}
\ee
\jump

5. Let us consider the effect of flavor-changing 
squark mixing.  
The mixing is induced by the Yukawa 
interactions in the superpotential 
and by the corresponding 
soft symmetry breaking terms. 
The mixing of the right handed components 
$\tilde b^c - \tilde s^c$ and   
$\tilde b^c - \tilde d^c$ 
proceeds via one loop diagrams formed by 
$u_p \tilde h$ or $\tilde u_p h,$ or by loops of $h$ and 
$\tilde u_p$  connected to 
$\tilde b^c - \tilde s^c$ by four boson coupling. 
Summation over $p$ leads to GIM cancellation which 
renders finite the contribution, and the 
suppression factor can be estimated as 
\be
\xi_{bs} \approx
\frac{1}{(4 \pi)^2} \times 
\frac{m_b m_s}{v^2} \ 
\frac{m_t^2}{\tilde{m}^2}\times
V_{23} \ V_{33}   
\approx 3 \cdot 10^{-8} \ .
\label{suppnon-m2}  
\ee
For $\tilde b^c - \tilde d^c$ mixing 
$m_s V_{23}$ 
should be substituted by 
$m_d V_{13}$. 

In the case of the left-right type mixing:    
$\tilde b^c - \tilde s$ and   
$\tilde b^c - \tilde d$ the $u_p$ quark in the loop 
should change chirality:   
$\tilde b^c \rightarrow 
t \tilde h \rightarrow t^c \tilde h \rightarrow  \tilde s$.  
There is no GIM cancellation and the diagrams are logarithmically 
divergent. A typical suppression factor equals    
\be
\xi_{bs} \approx  
\frac{1}{(4 \pi)^2} \times 
\frac{m_b m_t}{v^2}\ 
\frac{m_t m_{\tilde h}}{\tilde{m}^2}\times 
V_{23}\ V_{33}\times   
\ln\frac{M_{P}}{\tilde m}   
\approx 3 \cdot 10^{-6}, 
\label{suppnon-m3}  
\ee
where the renormalization point has been chosen at 
the Planck scale $M_P$. For 
$\tilde b^c - \tilde d$ mixing the suppression 
factor can be obtained 
from 
(\ref{suppnon-m3}) 
by the substitution: $V_{23} \rightarrow V_{13}$. 

The suppression factors
(\ref{suppnon-m2}, 
\ref{suppnon-m3}) are consistent with the bounds 
on the flavor-changing neutral currents, and there is no
reason to disregard flavor-changing squark mixing.

The flavor-changing mixing is important for 
restrictions on the couplings with large 
number of light generation indices. 
New tree level diagrams appear with squark mixing 
in the propagator. 
In fact, for all pairs of the couplings (\ref{best},\ref{next}) 
one can construct such a kind of diagram. 
This leads to suppression factors of 
the order $10^{-8}  - 10^{-6}$ instead of 
$10^{-15}$ found previously in absence of mixing. 
Correspondingly, one gets very strong bounds
$|\lambda' \cdot \lambda''| < 10^{-16}$. 

What are the largest allowed products of couplings 
in this case?  
Tree level diagrams with squark mixing for proton 
decay are absent for 
pairs of couplings with (1) two or more unmatching heavy fields  
(which can not be produced for kinematical reason)  or  
(2) with more than two $s$-fields. 
For example, the first criteria is satisfied for 
couplings  with $c^c$- or $t^c$-quark-fields instead of   
$u^c$. The smallest suppression factors  are found for the 
pairs  
\be 
\lambda'_{1j1}\ \lambda''_{232}, \ \
\lambda'_{2j1}\ \lambda''_{132}, \ \
\lambda'_{3j1}\ \lambda''_{122}, \ \
\lambda'_{3j3}\ \lambda''_{131}, \ \
\lambda'_{3j3}\ \lambda''_{231}, 
\label{fcincluded}
\ee    
where $j = 1,2,3$.  
The main contributions to the proton decay come from 
the following diagrams: 
pairs 1,2---``no-matching'' type vertex diagram 2b;
pair 3---``$B^c B$-matching'' vertex, diagram 1d;
pairs 4,5---``$B^c B$-matching'' box diagram of Fig.\ 1e.
The suppression factor 
$\xi \approx 10^{-13}$ leads to the bound on the 
products of couplings (\ref{fcincluded}): 
\be
|\lambda' \cdot \lambda''| \ltap  10^{-11} \ .
\label{bound}
\ee 

One remark is in order. 
The tree level diagram for the above couplings 
contains only one 
heavy quark $b$. In general due to the 
wave function renormalization this quark can be 
transformed into a $d$ or a $s$ leading to very 
fast proton decay. However such a renormalization 
is absent for our definition of the couplings: 
Recall that they are introduced in the basis 
where the fermionic components of the supermultiplets 
coincide with the mass states 
in any order of perturbation theory. 
This is equivalent to the redefinition of the 
couplings in each order of the perturbation 
theory.\jump 

6. Let us consider the dependence of the bounds 
on  basis in which the couplings are defined.  
As we marked before, the R-parity breaking couplings depend  
on the rotation matrices $S^u$ and $S^d$ separately, 
as well as on the possible rotation of the 
$D^c,U^c,L$ and $E^c$ superfields.
For example,  in the 
basis where $S^d = \V $ and $S^u =I$, 
there is ``matching'' for all pairs of couplings. 

In general, the bounds in any basis can be obtained from 
the bounds in the original basis by an appropriate rotation. 
Let us show that the conservative bound (\ref{bound}) 
is in fact basis independent. 

In a basis characterized by   
$S^d \neq I$  
the coupling constants $\lambda_{ijk}'$ 
are related to the constants in the original basis,  
$\lambda_{ijk}^{0}\!\!\!{}'\  $,  
as 
\be
\lambda_{ijl}'\ S_{lk}^d = 
\lambda_{ijk}^0\!\!\!{}'\ . 
\ee
Since $S_{lk}^d$ is  an  unitary transformation we get from the bound 
$|\lambda_{ijl}^0\!\!\!\;{}'\  | < {\cal B}_0$: 
\be
\sum_k |\lambda_{ijk}'|^2 = 
\sum_l |\lambda_{ijk}^0\!\!\!{}'\  |^2 <
3\ {\cal B}_0^2 , 
\ee
where ${\cal B}_0$ is the conservative bound on all modes in the 
original basis. 
Similarly one can estimate the effect of the
rotation of the other superfields.\jump  

7.  Final remarks. We have shown that for any pair 
of R-parity violating couplings it 
is possible to construct one loop diagrams 
which lead to the proton decay. 

If one disregards the 
flavor-changing squark mixing 
then the most conservative bound on the product of couplings is 
$
|\lambda'\cdot \lambda''|\ltap 10^{-9}
$
for squark masses below the TeV scale. 
If one takes into account
flavor-changing squark mixing then new diagrams appear and 
the conservative bound becomes stronger: 
$
|\lambda'\cdot \lambda''|\ltap 10^{-11} \ .
$

The pairs of coupling which have the weakest suppression 
are given in  \refe{best}, \refe{next} and \refe{fcincluded}.
It is interesting to observe that the couplings involved
are not those with 
maximal number of the heavy generation indices. 
In fact these couplings contain typically several light indices. 
Therefore it may be difficult to explain
the dominance of these couplings by some horizontal
symmetry, especially in models which reproduce
the hierarchy of the fermion masses. 
Let us admit however that some mechanism exists which picks up
a pair of constants which are weakly suppressed.
In this case the renormalization of these $\lambda$ due to
Yukawa interactions will induce R-parity violating couplings
which are not weakly bounded.
To solve this problem one should suggest that this mechanism 
operates at the scale $\mu\sim \tilde m \sim 1$ TeV.
This mechanism will not operate (or will imply further fine-tuning)
if $\mu$ is of the order of the Grand Unification scale
$M_{GU}$ or of the Planck scale.

The bounds \refe{conservative} and 
\refe{bound} are considerably weaker than 
the bound \refe{tree}, and weaker than the typical bound
which can be obtained  in the context of Grand Unified theories.
The bound \refe{conservative} is marginally 
compatible with both lepton {\em and} baryon violating effects which 
could be observable at accelerators,
$\lambda_{\rm obs}$ 
$\gtap 2\cdot 10^{-5}$  
$\sqrt{\gamma}$
$({\tilde m}/1\ {\rm TeV})^2$ 
$(150\ {\rm GeV}/ m_{\chi})^{5/2},$ 
where $\gamma$ is the Lorentz
boost factor \cite{p7}.
For example, a neutralino $\chi$ of approximatively 
150 GeV (or heavier) mass decays within the detector by
couplings $\lambda' \sim \lambda''\sim 2\cdot 10^{-5}$ which
satisfy the proton decay bound \refe{conservative}.
However, according to previous discussion, this case would require 
various fine-tunings of the model. 

The bounds discussed in the present paper will be strengthened at least 
by 1 order of magnitude if the SuperKamiokande experiment 
will not detect a signal of proton decay.
~~~~\vskip.5truecm~~~~
\noindent{\Large\bf Acknowledgements.}
\vskip0.6truecm

The authors  would like to thank  Z. Berezhiani, 
S. Bertolini, F. Borzumati, C. Savoy, G. Senjanovi\'c and M. Vysostsky 
for useful discussions.
\vfill\eject
\noindent{\Large\bf Figure Captions}
\vskip2truecm
\noindent{\large Fig.\ 1:}
{Feynman diagrams inducing proton decay in the
(anti)neutrino-meson(s) channels for the ``matching'' case, 
but not for the ``no-matching'' case. 
The blob indicates a fermion mass insertion.}
\vskip1truecm
\noindent{\large Fig.\ 2:}
{Feynman diagrams inducing proton decay in the
(anti)neutrino-meson(s) channels for the ``no-matching'' case. 
Compare with previous figure.}


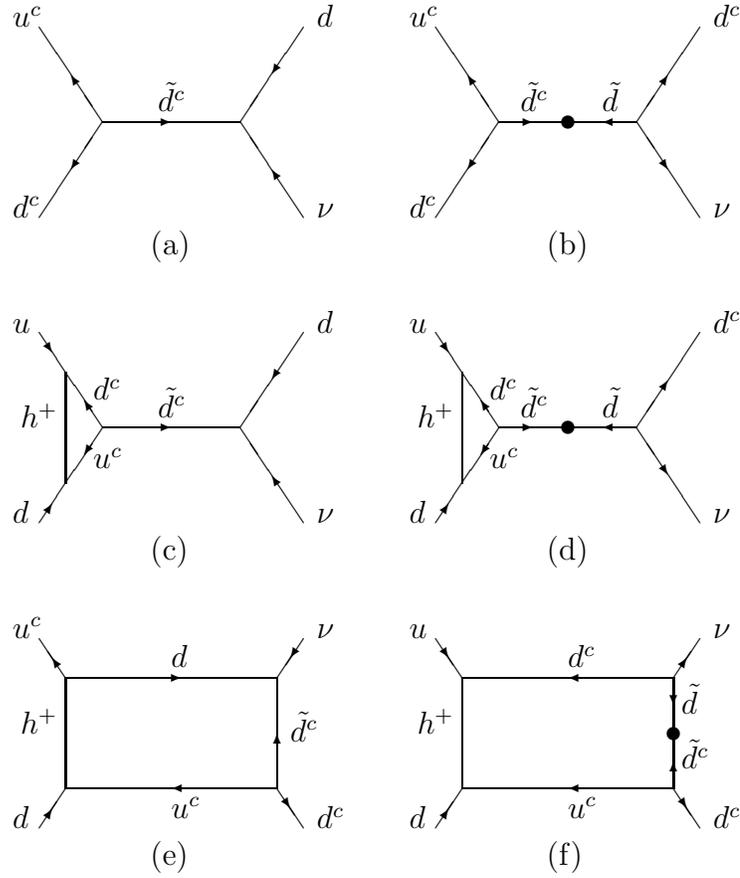
\begin{figure}[t]

\begin{center}

\begin{picture}(250,72)

\put(42,-14){(a)}
\put(192,-14){(b)}


\put(0,0){\line(2,3){24}}         
\put(12,18){\vector(-2,-3){0}}
\put(-10,0){$d^c$}

\put(0,72){\line(2,-3){24}}       
\put(12,54){\vector(-2,3){0}}
\put(-10,72){$u^c$}

\put(24,36){\line(1,0){52}}       
\put(50,36){\vector(1,0){0}}
\put(45,40){$\tilde{d}^c$}

\put(100,0){\line(-2,3){24}}      
\put(88,18){\vector(-2,3){0}}
\put(105,0){$\nu$}

\put(100,72){\line(-2,-3){24}}    
\put(88,54){\vector(-2,-3){0}}
\put(105,72){$d$}


\put(150,0){\line(2,3){24}}         
\put(162,18){\vector(-2,-3){0}}
\put(140,0){$d^c$}

\put(150,72){\line(2,-3){24}}       
\put(162,54){\vector(-2,3){0}}
\put(140,72){$u^c$}

\put(174,36){\line(1,0){52}}       
\put(187,36){\vector(1,0){0}}
\put(200,36){\circle*{5}}
\put(213,36){\vector(-1,0){0}}
\put(182,40){$\tilde{d}^c$}
\put(213,40){$\tilde{d}$}

\put(250,0){\line(-2,3){24}}      
\put(238,18){\vector(2,-3){0}}
\put(255,0){$\nu$}

\put(250,72){\line(-2,-3){24}}    
\put(238,54){\vector(2,3){0}}
\put(255,72){$d^c$}

\end{picture}

\vskip1.5truecm

\begin{picture}(250,72)

\put(42,-14){(c)}
\put(192,-14){(d)}


\put(10,15){\line(0,1){42}}       
\put(-7,36){$h^{ +}$}

\put(0,0){\line(2,3){24}}         
\put(5,7.5){\vector(2,3){0}}
\put(-10,0){$d$}

\put(0,72){\line(2,-3){24}}       
\put(5,64.5){\vector(2,-3){0}}
\put(-10,72){$u$}

\put(17,25.5){\vector(-2,-3){0}}    
\put(20.5,20.5){$u^c$}

\put(17,46.5){\vector(-2,3){0}}    
\put(20.5,46.5){$d^c$}

\put(24,36){\line(1,0){52}}       
\put(50,36){\vector(1,0){0}}
\put(45,40){$\tilde{d}^c$}

\put(100,0){\line(-2,3){24}}      
\put(88,18){\vector(-2,3){0}}
\put(105,0){$\nu$}

\put(100,72){\line(-2,-3){24}}    
\put(88,54){\vector(-2,-3){0}}
\put(105,72){$d$}


\put(160,15){\line(0,1){42}}       
\put(143,36){$h^{ +}$}

\put(150,0){\line(2,3){24}}         
\put(155,7.5){\vector(2,3){0}}
\put(140,0){$d$}

\put(150,72){\line(2,-3){24}}       
\put(155,64.5){\vector(2,-3){0}}
\put(140,72){$u$}

\put(167,25.5){\vector(-2,-3){0}}    
\put(170.5,20.5){$u^c$}

\put(167,46.5){\vector(-2,3){0}}    
\put(170.5,46.5){$d^c$}

\put(174,36){\line(1,0){52}}       
\put(187,36){\vector(1,0){0}}
\put(200,36){\circle*{5}}
\put(213,36){\vector(-1,0){0}}
\put(182,40){$\tilde{d}^c$}
\put(213,40){$\tilde{d}$}

\put(250,0){\line(-2,3){24}}      
\put(238,18){\vector(2,-3){0}}
\put(255,0){$\nu$}

\put(250,72){\line(-2,-3){24}}    
\put(238,54){\vector(2,3){0}}
\put(255,72){$d^c$}

\end{picture}

\vskip1.5truecm

\begin{picture}(250,72)

\put(42,-14){(e)}
\put(192,-14){(f)}


\put(10,15){\line(0,1){42}}       
\put(-7,36){$h^{ +}$}

\put(0,0){\line(2,3){10}}         
\put(5,7.5){\vector(2,3){0}}
\put(-10,0){$d$}

\put(0,72){\line(2,-3){10}}       
\put(5,64.5){\vector(-2,3){0}}
\put(-10,72){$u^c$}

\put(10,15){\line(1,0){80}}       
\put(50,15){\vector(-1,0){0}}
\put(50,4){${u}^c$}

\put(10,57){\line(1,0){80}}       
\put(54,57){\vector(1,0){0}}
\put(50,61){${d}$}

\put(90,15){\line(0,1){42}}       
\put(90,36){\vector(0,1){0}}
\put(95,33){$\tilde{d}^c$}

\put(100,0){\line(-2,3){10}}      
\put(95,7.5){\vector(2,-3){0}}
\put(105,0){$d^c$}

\put(100,72){\line(-2,-3){10}}    
\put(95,64.5){\vector(-2,-3){0}}
\put(105,72){$\nu $}


\put(160,15){\line(0,1){42}}       
\put(143,36){$h^{ +}$}

\put(150,0){\line(2,3){10}}         
\put(155,7.5){\vector(2,3){0}}
\put(140,0){$d$}

\put(150,72){\line(2,-3){10}}       
\put(155,64.5){\vector(2,-3){0}}
\put(140,72){$u$}

\put(160,15){\line(1,0){80}}       
\put(200,15){\vector(-1,0){0}}
\put(200,4){${u}^c$}

\put(160,57){\line(1,0){80}}       
\put(200,57){\vector(-1,0){0}}
\put(200,61){${d}^c$}

\put(240,15){\line(0,1){42}}       
\put(240,25.5){\vector(0,1){0}}
\put(243,22.5){$\tilde{d}^c$}
\put(240,36){\circle*{5}}
\put(240,46.5){\vector(0,-1){0}}
\put(243,43.5){$\tilde{d}$}

\put(250,0){\line(-2,3){10}}      
\put(245,7.5){\vector(2,-3){0}}
\put(255,0){$d^c$}

\put(250,72){\line(-2,-3){10}}    
\put(245,64.5){\vector(2,3){0}}
\put(255,72){$\nu $}

\end{picture}

\end{center}

\vskip0.7truecm

\caption{Feynman diagrams inducing proton decay in the
(anti)neutrino-meson(s) channels for the ``matching'' case, 
but not for the ``no-matching'' case. 
The blob indicates a fermion mass insertion.
}

\end{figure}


\vfill\eject

\begin{figure}[t]
\begin{center}

\begin{picture}(260,120)

\put(42,5){(a)}
\put(192,5){(b)}

\put(0,15){$\nu$}
\put(10,15){\line(2,3){10}}             
\put(15,22.5){\vector(-2,-3){0}}

\put(0,105){$u$}
\put(10,105){\line(2,-3){10}}          
\put(15,97.5){\vector(2,-3){0}}

\put(20,30){\line(2,3){20}}           
\put(33,42){$\tilde d$}
\put(30,45){\vector(2,3){0}} 

\put(28,78){$h^+$}
\put(40,60){\line(-2,3){20}}          

\put(20,30){\line(0,1){60}}           
\put(10,57){$d^c$}
\put(20,60){\vector(0,1){0}}

\put(48,64){$\tilde{u}^c$}
\put(40,60){\line(1,0){20}}           
\put(50,60){\vector(-1,0){0}}

\put(95,15){$d^c$}
\put(90,15){\line(-2,3){30}}          
\put(75,37.5){\vector(2,-3){0}}

\put(95,105){$d^c$}
\put(90,105){\line(-2,-3){30}}       
\put(75,82.5){\vector(2,3){0}}

\put(150,15){$d^c$}
\put(160,15){\line(2,3){10}}             
\put(165,22.5){\vector(-2,-3){0}}

\put(150,105){$u$}
\put(160,105){\line(2,-3){10}}          
\put(165,97.5){\vector(2,-3){0}}

\put(170,30){\line(2,3){20}}           
\put(183,42){$\tilde u^c$}
\put(180,45){\vector(2,3){0}} 

\put(178,78){$h^+$}
\put(190,60){\line(-2,3){20}}          

\put(170,30){\line(0,1){60}}           
\put(160,57){$d^c$}
\put(170,60){\vector(0,1){0}}           

\put(198,64){$\tilde{d}$}
\put(190,60){\line(1,0){20}}           
\put(200,60){\vector(-1,0){0}}

\put(245,15){$\nu$}
\put(240,15){\line(-2,3){30}}          
\put(225,37.5){\vector(2,-3){0}}

\put(245,105){$d^c$}
\put(240,105){\line(-2,-3){30}}       
\put(225,82.5){\vector(2,3){0}}

\end{picture}

\vskip0truecm

\begin{picture}(260,120)

\put(42,5){(c)}
\put(192,5){(d)}

\put(0,15){$\nu$}
\put(10,15){\line(2,3){10}}             
\put(15,22.5){\vector(-2,-3){0}}

\put(0,105){$u^c$}
\put(10,105){\line(2,-3){10}}          
\put(15,97.5){\vector(-2,3){0}}

\put(20,30){\line(2,3){20}}           
\put(33,42){$\tilde d$}
\put(30,45){\vector(2,3){0}} 

\put(28,78){$h^+$}
\put(40,60){\line(-2,3){20}}          

\put(20,30){\line(0,1){60}}           
\put(10,42){$d^c$}
\put(20,45){\vector(0,1){0}}           
\put(10,75){$d$}
\put(20,75){\vector(0,-1){0}}           

\put(20,60){\circle*{5}}           

\put(48,64){$\tilde{u}^c$}
\put(40,60){\line(1,0){20}}           
\put(50,60){\vector(-1,0){0}}

\put(95,15){$d^c$}
\put(90,15){\line(-2,3){30}}          
\put(75,37.5){\vector(2,-3){0}}

\put(95,105){$d^c$}
\put(90,105){\line(-2,-3){30}}       
\put(75,82.5){\vector(2,3){0}}

\put(150,15){$d^c$}
\put(160,15){\line(2,3){10}}             
\put(165,22.5){\vector(-2,-3){0}}

\put(150,105){$u^c$}
\put(160,105){\line(2,-3){10}}          
\put(165,97.5){\vector(-2,3){0}}

\put(170,30){\line(2,3){20}}           
\put(183,42){$\tilde u^c$}
\put(180,45){\vector(2,3){0}} 

\put(178,78){$h^+$}
\put(190,60){\line(-2,3){20}}          

\put(170,30){\line(0,1){60}}           
\put(160,42){$d^c$}
\put(170,45){\vector(0,1){0}}           
\put(160,75){$d$}
\put(170,75){\vector(0,-1){0}}           

\put(170,60){\circle*{5}}           

\put(198,64){$\tilde{d}$}
\put(190,60){\line(1,0){20}}           
\put(200,60){\vector(-1,0){0}}

\put(245,15){$\nu$}
\put(240,15){\line(-2,3){30}}          
\put(225,37.5){\vector(2,-3){0}}

\put(245,105){$d^c$}
\put(240,105){\line(-2,-3){30}}       
\put(225,82.5){\vector(2,3){0}}

\end{picture}

\vskip0truecm

\begin{picture}(260,120)

\put(42,5){(e)}
\put(192,5){(f)}

\put(0,15){$\nu$}
\put(10,15){\line(2,3){10}}             
\put(15,22.5){\vector(2,3){0}}

\put(0,105){$u^c$}
\put(10,105){\line(2,-3){10}}          
\put(15,97.5){\vector(-2,3){0}}

\put(20,30){\line(2,3){20}}           
\put(27,30){$\tilde{d}^c$}
\put(23,34.5){\vector(-2,-3){0}}
\put(37,45){$\tilde{d}$}
\put(38,57){\vector(2,3){0}} 

\put(30,45){\circle*{5}}            

\put(28,78){$h^+$}
\put(40,60){\line(-2,3){20}}          

\put(10,55){$d$}
\put(20,30){\line(0,1){60}}           
\put(20,60){\vector(0,-1){0}}           

\put(48,64){$\tilde{u}^c$}
\put(40,60){\line(1,0){20}}           
\put(50,60){\vector(-1,0){0}}

\put(95,15){$d^c$}
\put(90,15){\line(-2,3){30}}          
\put(75,37.5){\vector(2,-3){0}}

\put(95,105){$d^c$}
\put(90,105){\line(-2,-3){30}}       
\put(75,82.5){\vector(2,3){0}}

\put(150,15){$d^c$}
\put(160,15){\line(2,3){10}}             
\put(165,22.5){\vector(-2,-3){0}}

\put(150,105){$u$}
\put(160,105){\line(2,-3){10}}          
\put(165,97.5){\vector(2,-3){0}}

\put(170,30){\line(2,3){20}}           
\put(177,30){$\tilde{u}^c$}
\put(176.6666666,40){\vector(2,3){0}}
\put(187,45){$\tilde{u}$}
\put(183.3333333,50){\vector(-2,-3){0}}

\put(180,45){\circle*{5}}            

\put(178,78){$h^+$}
\put(190,60){\line(-2,3){20}}          

\put(160,55){$d^c$}
\put(170,30){\line(0,1){60}}           
\put(170,60){\vector(0,1){0}}           

\put(198,64){$\tilde{d}^c$}
\put(190,60){\line(1,0){20}}           
\put(200,60){\vector(1,0){0}}

\put(245,15){$\nu$}
\put(240,15){\line(-2,3){30}}          
\put(225,37.5){\vector(-2,3){0}}

\put(245,105){$d$}
\put(240,105){\line(-2,-3){30}}       
\put(225,82.5){\vector(-2,-3){0}}

\end{picture}

\end{center}

\caption{Feynman diagrams inducing proton decay in the
(anti)neutrino-meson(s) channels for the ``no-matching'' case. 
Compare with previous figure.}

\end{figure}
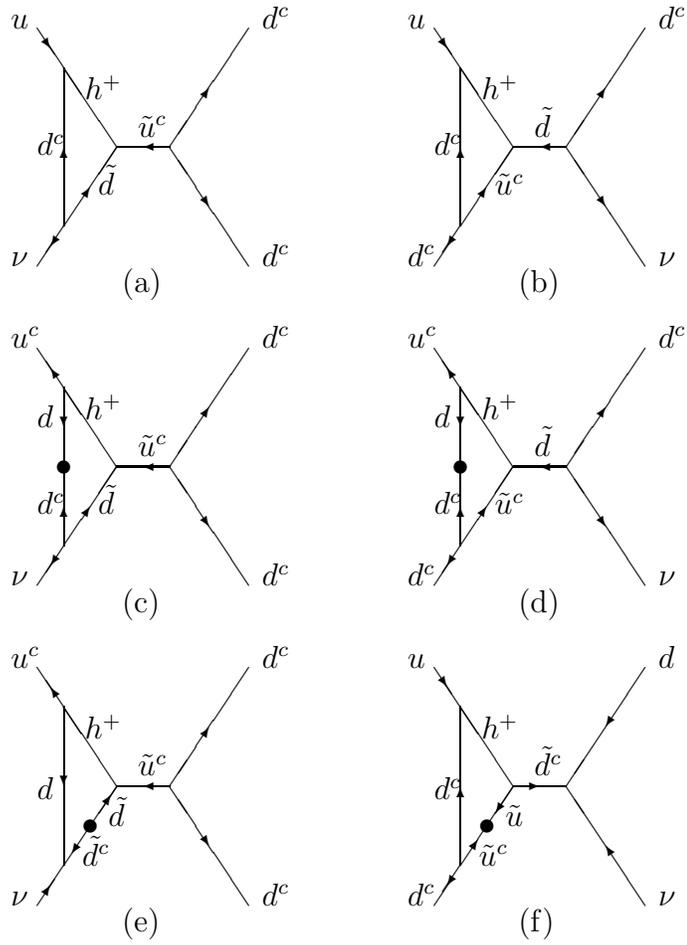


\end{document}